\documentclass[a4paper]{spie}  
\usepackage[dvips]{graphicx}

\def\CN2{\mbox{$C_N^2 \ $}}

\def\CT2{\mbox{$C_T^2 \ $}}

\def\sigmal2{\mbox{$\sigma ^{2}_{I} \ $}}

\title{Mt. Graham: Optical turbulence vertical distribution at standard and high vertical resolution} 

\author{Elena Masciadri\supit{a}, Jeff Stoesz\supit{a}, Susanna Hagelin\supit{a,b}, Franck Lascaux\supit{a}
\skiplinehalf
\supit{a}INAF - Osservatorio Astrofisico di Arcetri, L.go E. Fermi 5, 50125  Florence, Italy
\skiplinehalf
\supit{b}Departement of Earth Sciences, Uppsala University, Uppsala, Sweden\\
}

\authorinfo{Correspondence to Elena Masciadri: masciadri@arcetri.astro.it}

\begin{document} 
  \maketitle 

\begin{abstract}
A characterization of the optical turbulence vertical distribution and all the main integrated astroclimatic parameters derived from the \CN2 and the wind speed profiles above Mt. Graham is presented. The statistic includes measurements related to 43 nights done with a Generalized Scidar (GS) used in standard configuration with a vertical resolution of $\sim$1 km on the whole 20-22 km and with the new technique (HVR-GS) in the first kilometer. The latter achieves a resolution of $\sim$ 20-30 m in this region of the atmosphere. Measurements done in different periods of the year permit us to provide a seasonal variation analysis of the \CN2. A discretized distribution of the typical \CN2 profiles useful for the Ground Layer Adaptive Optics (GLAO) simulations is provided and a specific analysis for the LBT Laser Guide Star system ARGOS case is done including the calculation of the 'gray zones' for J, H and K bands. Mt. Graham confirms to be an excellent site with median values of the seeing without dome contribution equal to 0.72", the isoplanatic angle equal to 2.5" and the wavefront coherence time equal to 4.8 msec. We provide a cumulative distribution of the percentage of turbulence developed below H* where H* is included in the (0,1 km) range.  We find that 50\% of the whole turbulence develops in the first 80 m from the ground. The turbulence decreasing rate is very similar to what has been observed above Mauna Kea.
\end{abstract}



\section{INTRODUCTION}
\label{sec:intro} 
The Mt. Graham International Observatory (MGIO) is located on Mt. Graham (32$^{\circ}$42'05" N, 109$^{\circ}$53'31" W), Arizona (US) and hosts three telescopes: the Vatican Advanced Technological Telescope (VATT - D = 1.83 m) the Heinrich Hertz Submillimiter Telescope (SMT - D = 10 m) and the Large Binocular Telescope (LBT - two D = 8.4 m dishes that, when working in an  interferometric configuration, can achieve the resolution of a telescope with a 23 m pupil size). The study and characterization of the optical turbulence (OT) distribution in space and time is fundamental for the ground-based astronomy in the visible up to the near-infrared range to design adaptive optics systems and to optimize their performances. The vertical distribution of the OT (i.e. the $\CN2$ profiles) is the parameter from which all the integrated astroclimatic parameters derive from. 

In this paper we present a study based on measurements of the $\CN2$ profiles related to 43 nights and obtained with a Generalized Scidar (GS) placed at the focus of the VATT on the Mt.Graham summit, around 250 m from the LBT. We invite the readers to refer to the paper Masciadri et al. (2010)\cite{Masciadri2010} for a more extensively and complete analysis. In this context, we will try to summarize the most important results obtained and to focus on the applications to the Adaptive Optics applications.

The scientific motivations for such a long-term site testing campaign are: \newline
{\bf (1)} To collect an as rich as possible statistical sample of optical turbulence (OT) vertical distribution ($\CN2$ profiles) to be compared with simulations obtained with the atmospherical mesoscale model Meso-Nh with the aim to validate the model above Mt. Graham. This is a key milestone for the ForOT project\footnote{$http://forot.arcetri.astro.it$} whose final goal is to predict the optical turbulence above astronomical sites\cite{Masciadri2006}. The measurements from a vertical profiler such as a Generalized Scidar are crucial for the validation of such a kind of models. It is our interest to collect a heterogeneous sample of measurements taken in different periods of the year and different turbulence conditions in a way to better control the model behaviors under different conditions and to better validate the model itself. The atmospherical models has been used for the first time to reconstruct and characterize the $\CN2$ profiles by Masciadri et al.\cite{Masciadri1999a,Masciadri1999b}. Since that time many progresses have been achieved by our group: the model has been applied to different astronomical sites in a simple monomodel configuration\cite{Masciadri2001,MasciadriGarfias2001} and, more recently, in a grid-nesting configuration\cite{Lascaux2009,Lascaux2010}. A new calibration technique has been proposed\cite{MasciadriJabouille2001} and statistically validated\cite{Masciadri2004} and the first application of the Meso-Nh model as a tool of turbulence characterization has been presented\cite{MasciadriEgner2006}. However, so far we always could access to GS measurements concentrated in a well defined period of time. The site testing campaigns done with a GS at Mt. Graham aimed to comply with the necessity to diversify the measurements sample. \newline
{\bf (2)} To provide a characterization of all the most important integrated astroclimatic parameters above the site of the Large Binocular Telescope (LBT). \newline 
{\bf (3)} To provide an as rich as possible statistical sample of the high resolution vertical distribution ($\Delta$H = 200-250 m and 25-30 m) of the optical turbulence in the first hundreds of meters up to 1 km to support the feasibility studies of new generation instruments for the LBT, such as the LBT Laser Guide Stars system ARGOS\cite{Rabien2008,Rabien2010}, that, in its first baseline, is planned to work with a GLAO\footnote{Ground Layer Adaptive Optics (GLAO)} configuration. The GLAO efficiency, indeed, strongly depends on the turbulence distribution and strength near the ground. The study we intend to perform can be achieved thanks to a new technique that has been recently proposed called High-Vertical Resolution Generalized Scidar (HVR-GS\cite{EgnerMasciadri2007}) that aims to reconstruct the optical turbulence vertical distribution in the first kilometer above the ground with a resolution 4 up to 10 times higher than what has been done so far with standard vertical profilers such as the GS (typically $\Delta$H $\sim$ 1km). In Egner \& Masciadri (2007)\cite{EgnerMasciadri2007} the validity of this technique has been proved and it is now our intention to characterize the turbulence distribution in statistical terms, to verify how the turbulence decreases in the first hundreds of meters above the ground and provide inputs to test the ARGOS performances.

\section{INSTRUMENTS AND DATA-SETS}
Two instruments have been used for this study: the GS and the HVR-GS. We used the GS as developed by McKenna et al.\cite{McKenna2003}. The HVR-GS technique has been introduced recently\cite{EgnerMasciadri2007} and aims to measure the $\CN2$ profiles with a high vertical resolution ($\Delta$H $\sim$ 25-30 m) in the first kilometer above the ground. We refer to Egner \& Masciadri (2007)\cite{EgnerMasciadri2007} for the details of the techniques. 

\begin{table}
\caption{Classification of the GS campaign measurements. {\bf 'GS':} $\CN2$ profiles retrieved from the standard GS.  {\bf 'WB':} $\CN2$ profiles retrieved from the AC frames of the GS measurements obtained with wide-binaries. {\bf 'HVR-GS':} $\CN2$ profiles retrieved from the wide-binaries auto-correlation (AC) and cross-correlation (CC) frames following the technique described in Egner \& Masciadri (2007)\cite{EgnerMasciadri2007}.}
\begin{center}
{\begin{tabular}{ccccc}
\hline
Sample & Nights & Measurements & Hours & Resolution\\
\hline
'GS' & 43 & 16657 &  163 &  $\Delta$H(0) $\sim$ 1km\\
'WB' & 15 & 3659 &  6.2 &  $\Delta$H(0) $\sim$ 200 m\\
'HVR-GS' & 15 & 2812 &  5.1 &  $\Delta$H(0) $\sim$ 25 m\\
\hline
\end{tabular}}
\end{center}
\label{tab_sample}
\end{table}

We collected and analyzed observations related to 43 nights (Table 1 in Masciadri et al. (2010)\cite{Masciadri2010}). Table \ref{tab_sample} shows the code used to identify the typology of the data-set: 'GS' indicates the standard GS measurements extended on $\sim$ 20 km (43 nights), 'WB' indicates the $\CN2$ retrieved from the autocorrelation (AC) frames associated to wide binaries ($\Delta$H $\sim$ 200-250 m) (15 nights) and 'HVR-GS' the $\CN2$ obtained with high vertical resolution ($\Delta$H $\sim$ 25-30 m) (15 nights) in which both the autocorrelation and cross-correlation frames have been treated. The samples of the three categories are differently rich because we started to use the HVR-GS more recently and this new technique has a higher rate of measurements rejection. We note that, to avoid biases in the estimates in the HVR-GS, it has been decided to discard from the statistic all doubtful cases characterized by the presence of clouds or cirrus. A method that we called {\it 'normalization'}\cite{Masciadri2010} has been applied to the sample of high resolution measurements (15 nights). Among other advantages it permits us to provide a turbulence budget representative of the whole sample of 43 nights. Basically, the morphology of the turbulence energy distribution (shape of the $\CN2$ versus the height) is retrieved from the observation of wide-binaries for 15 nights and with the {\it 'normalization'} procedure the turbulence energy of the first kilometer detected with the standard GS is redistributed in thin turbulent layers according to the profile morphology reconstructed with the CC frames.

\section{INTEGRATED ASTROCLIMATIC PARAMETERS}
\label{astro_integ}

Table \ref{tab_integ_astro} summarizes the median (50$^{th}$), first (25$^{th}$) and third (75$^{th}$) quartiles for the three main integrated astroclimatic parameters calculated for the following groups: the whole sample of 43 nights, the summer and the winter time. 
A composite wind speed profile has been used to calculate the median wavefront coherence time $\tau_{0}$. Below 2 km the wind speed retrieved from the GS has been used; above 2 km the wind speed profile retrieved from the European Centre for Medium-Range Weather Forecast (ECMWF) analyses extracted in the nearest grid point (32.75$^{\circ}$N, 110.00$^{\circ}$W) to the Mt. Graham ($\sim$ 11.5 km northwest of the summit) has been used. The composite method for the wind speed revealed to be the best solution for the calculation of $\tau_{0}$\cite{Egner2007}. We refer the readers to Masciadri et al. (2010)\cite{Masciadri2010} for the calculation of the cumulative distribution of all the integrated astroclimatic parameters. 
Finally, to quantify the contribution of the seeing provided only by the atmosphere, the dome seeing ($\varepsilon_{d}$), calculated with the method described by Avila et al. (2001)\cite{Avila2001}, has been subtracted from the total seeing ($\varepsilon$). Knowing that the median seeing in the whole atmosphere (included the dome seeing) is $\varepsilon$$=$0.95 arcsec and that $\varepsilon_{d}$$=$ 0.52 arcsec, it follows that the median seeing related to the whole atmosphere without the dome contribution for the richest statistic we collected so far (43 nights) is $\varepsilon_{tot}$$=$ 0.72 arcsec.

\begin{table*}
\caption{Median, first and third quartiles values of the main integrated astroclimatic parameters above Mt. Graham (43 nights): seeing in the total troposphere, isoplanatic angle, wavefront coherence time, integrated equivalent wind speed.}
\begin{center}
{\begin{tabular}{cccccccccc}
\hline
&Total &&& Summer & && Winter  && \\
\hline
Parameter & 25$^{th}$ & 50$^{th}$ & 75$^{th}$& 25$^{th}$ & 50$^{th}$ & 75$^{th}$& 25$^{th}$ & 50$^{th}$ & 75$^{th}$\\
\hline
$\varepsilon$ (arcsec)& 0.65 & 0.95 & 1.34 &0.53 &0.61 &0.72 &0.89 &1.19 &1.50  \\
$\theta_{0}$ (arcsec) & 1.6  & 2.5  &  3.6 &3.1 & 3.8 & 4.5& 1.4& 2.0& 2.7 \\
$\tau_{0}$ (msec) & 2.7 & 4.8 & 8.7& 6.4& 10.1&14.6 &2.5 &3.8 & 6.2 \\
V$_{0}$ (ms$^{-1}$) & 5.1 & 7.2& 9.3 &3.7 &5.1 &7.8 &5.9 &7.7 & 9.6 \\
\hline
\end{tabular}}
\end{center}
\label{tab_integ_astro}
\end{table*}

\section{COMPOSITE PROFILES FOR ADAPTIVE OPTICS}
\label{composit}

The {\it 'composite profiles'} are used to represent efficiently and in statistical terms the vertical distribution of the optical turbulence in a discretized number of layers particularly suitable for AO simulations. The method is commonly used by many authors in the field\cite{TokovininTravouillon2006,Egner2007,Stoesz2008,Masciadri2010,Chun2009} and it consists in identifying a finite number of vertical slabs covering the whole troposphere ($\sim$ 20 km) with their correspondent value of seeing ($\Delta$$\varepsilon$$_{i}$ or $\Delta$J$_{i}$) so that the total turbulence integrated on the troposphere is conserved. J is defined as:

\begin{equation}
J = \int\limits_0^\infty  {C_N^2 (h) dh} 
\end{equation}

and it is related to the seeing as:

\begin{equation}
J = 9 \cdot 10^{-11} \cdot \lambda ^{1/3}  \cdot \varepsilon ^{5/3} 
\end{equation}

where $\lambda$ is expressed in meters, $\varepsilon$ in arcsec and J in m$^{1/3}$. The turbulence in the free atmosphere (h $>$ 1 km) and in the boundary layer (h $\le$ 1 km) is treated in an independent way to permit to study different combinations of probabilities for the OT vertical distribution. 
Table \ref{comp_h_gt_1km} reports the $\Delta$J$_{i}$ values in the range h $>$ 1 km calculated at different heights and obtained from the $\CN2$ associated to the r$_{0}$ related to the 20-30 $\%$ of its cumulative distribution ('good' case), to the 45-55 $\%$ ('typical' case) and to the 70-80 $\%$ ('bad' case). Measurements from the sample 'GS' (see Table \ref{tab_sample}) are used. 

Again we refer the reader to Masciadri et al. (2010)\cite{Masciadri2010} for precise information on data-reduction analysis. The $\CN2$ profiles retrieved from the 'WB' sample characterized by a $\Delta$h $\sim$ 200 m near the ground (H $<$ 1 km) have a suitable vertical resolution to calculate the composite profiles in this vertical range for applications to ARGOS (field of view $\theta$ = 4 arcmin) because the turbulence developed below H$_{min}$ = $\Delta$X/(2$\cdot$$\theta$) $\sim$ 200 m ($\Delta$X = 0.5 m is the pitch size i.e. the projection of the actuator of the deformable mirror on the pupil of the telescope) is resolved by the instrument. Basically we do not need a higher vertical resolution for this application. 
Table \ref{comp_h_lt_1km_corr} reports the composite distribution for h $\le$ 1 km equivalent to 43 nights. It includes the dome contribution frequently preferable for AO simulations. 
The median value of the composite profiles (Table \ref{comp_h_gt_1km} and Table \ref{comp_h_lt_1km_corr} central columns) permits us to calculate the percentage of the turbulence developed above different heights h with respect to the turbulence developed in the whole troposphere (Table \ref{J}-left side) as well as the percentage of turbulence developed in the (0, $h$) range with respect to the turbulence developed in the whole troposphere (Table \ref{J}-right side). Table \ref{Jd} shows the same calculation obtained in case the median dome seeing is subtracted.

\begin{table}
\caption{Composite profiles for h $>$ 1 km. In the first column the boundaries of the vertical slabs. In each column is reported the value of J that is proportional to the integral of the optical turbulence in the correspondent vertical slab. These composite profiles are statistically representative for 43 nights.}
\begin{center}
{\begin{tabular}{cccc}
\hline
Bins &'Good'  &'Typical' &'Bad'  \\
(m) &  J (m$^{1/3}$) & J (m$^{1/3}$) & J (m$^{1/3}$) \\
\hline
14000-20000 & 7.61e-15 &  1.02e-14 &  1.91e-14 \\
12000-14000 & 5.10e-15 &  9.65e-15 &  1.57e-14 \\
10000-12000 & 7.24e-15 &  1.27e-14 &  2.89e-14 \\
8000-10000 &  8.52e-15 &  1.72e-14 &  3.22e-14 \\
6000-8000 & 6.45e-15 &  1.05e-14 &  1.77e-14 \\
4000-6000 &  9.50e-15 &  1.23e-14 &  2.10e-14 \\
3000-4000 &  9.18e-15 &  1.14e-14 &  1.45e-14 \\
2000-3000 &  1.97e-14 &  3.39e-14 &  4.02e-14 \\
1500-2000 & 6.98e-15 &  1.52e-14 &  2.86e-14 \\
1000-1500 &  5.28e-15 &  1.50e-14 &  2.86e-14 \\
\hline
\end{tabular}}
\end{center}
\label{comp_h_gt_1km}
\end{table}

\begin{table}
\caption{Composite profiles for h $<$ 1 km after the {\it 'normalization'} for the f$_{gl}$ factor. These composite profiles are statistically representative for 43 nights. In the last line are reported the J values obtained without the dome contribution (median values: $\varepsilon_{d,25}$ = 0.35 arcsec, $\varepsilon_{d,50}$ = 0.52 arcsec, $\varepsilon_{d,75}$ = 0.70 arcsec).}
\begin{center}
{\begin{tabular}{cccc}
\hline
 Bins & 'Good'&  'Typical'& 'Bad'  \\
(m) & J (m$^{1/3}$) &  J (m$^{1/3}$) & J (m$^{1/3}$) \\
\hline
900-1000 & 4.35e-15 &  7.06e-15 &  1.33e-14 \\
800-900 & 2.48e-15 &  3.53e-15 &  6.29e-15 \\
700-800 & 3.30e-15 &  6.04e-15 &  1.04e-14 \\
600-700 & 5.72e-15 &  1.09e-14 &  2.05e-14 \\
500-600 & 4.30e-15 & 7.70e-15 &  1.38e-14 \\
400-500 & 4.21e-15 &  9.60e-15 &  1.86e-14 \\
300-400 &  2.05e-14 &  4.26e-14 &  8.26e-14 \\
200-300 & 6.15e-15 &  1.14e-14 &  2.53e-14 \\
100-200 & 2.03e-14 &  4.14e-14 & 8.11e-14 \\
0-100 & 1.79e-13 &  3.27e-13 &  6.17e-13 \\
\hline
0-100 &  3.4e-14 &  8.5e-14 &  2.20e-13 \\
\hline
\end{tabular}}
\end{center}
\label{comp_h_lt_1km_corr}
\end{table}

\begin{table}
\caption{{\bf Dome seeing included -} {\bf Left:} Percentage of turbulence developed above the height $h$ with respect to the turbulence developed on the whole troposphere. {\bf Right:} Percentage of turbulence developed between the ground and the height h with respect to the turbulence developed in the whole turbulence. Second and fourth columns are obviously complementary.}
\begin{center}
{\begin{tabular}{cccccc}
\hline
 h & J$_{[h,top]}$/J$_{tot}$ &&  h & J$_{[0,h]}$/J$_{tot}$\\
 (m)  &  (\%)  &&  (m) & (\%)\\
\hline
1000  & 25  && 1000 & 75 \\
800  & 26 &&   800&  74\\
600  & 29  && 600 & 71\\
400  & 31 &&  400& 69\\
300  & 38  && 300 & 62\\
200  & 40  &&  200& 60\\
100  & 47 &&  100 & 53\\
\hline
\end{tabular}}
\end{center}
\label{J}
\end{table}

\begin{table}
\caption{{\bf Dome seeing subtracted -} {\bf Left:} Percentage of turbulence developed above the height $h$ with respect to the turbulence developed on the whole troposphere. {\bf Right:} Percentage of turbulence developed between the ground and the height h with respect to the turbulence developed in the whole turbulence. Second and fourth columns are obviously complementary.}
\begin{center}
{\begin{tabular}{cccccc}
\hline
 h & J$_{[h,top]}$/J$_{tot}$ &&  h & J$_{[0,h]}$/J$_{tot}$\\
 (m)  &  (\%)  &&  (m) & (\%)\\
\hline
1000  & 40  && 1000 & 60 \\
800  & 43 &&   800&  57\\
600  & 47  && 600 & 53\\
400  &52 &&  400& 48\\
300  & 63  && 300 & 37\\
200  & 66  &&  200& 34\\
100  & 77 &&  100 & 23\\
\hline
\end{tabular}}
\end{center}
\label{Jd}
\end{table}

\begin{figure}
\centering
\includegraphics[width=7.5cm]{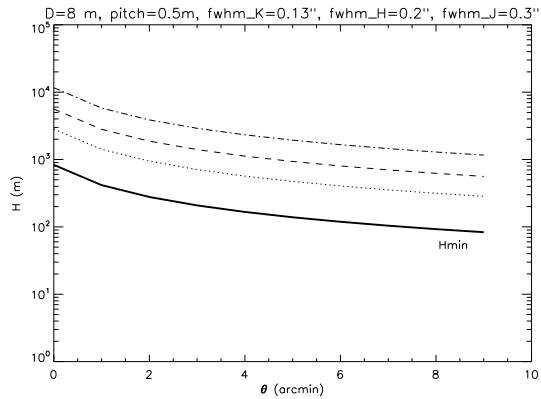}
\caption{Extent of the {\it 'gray zone'} i.e. H$_{min}$ $<$ h $<$ H$_{max}$ for different fields of view and wavelengths in the case of median distribution (50$\%$). H$_{max}$ is calculated for J (dotted line), H (dashed line) and K (dot-dashed-dot line) band. H$_{min}$ (bold style line) is the same for all the wavelengths. The pupil size is D = 8 m and the pitch size $\Delta$X = 0.5 m. The FWHM is equivalent to r i.e. the residual wavefront coherence size after correction.
\label{glao}} 
\end{figure}

\begin{table*}
\caption{Values of H$_{max}$ calculated for $\theta$ = 4 arcmin and different residual FWHM. The FWHM values are obtained with the GLAO simulations using as inputs Table \ref{comp_h_gt_1km} and Table \ref{comp_h_lt_1km_corr}. 
H$_{min}$ = $\Delta$X/2$\theta$ $\sim$ 200 m for all the wavelengths.}
\begin{center}
{\begin{tabular}{cccccccccc}
\hline
75$\%$  & FWHM& H$_{max}$ &50$\%$  & FWHM& H$_{max}$ &  25$\%$ & FWHM & H$_{max}$ \\
  & (arcsec) & (m) &  & (arcsec) & (m) &  & (arcsec)& (m) \\
\hline
J & 0.43& 378 &J & 0.30& 567 & J  & 0.18 & 945 \\
H & 0.37&606&H & 0.20&1123 &  H& 0.11&2042 \\
K & 0.25&1208 &K & 0.13&2324 & K & 0.08  &3777 \\
\hline
\end{tabular}}
\end{center}
\label{glao_tab}
\end{table*}

The composite distribution of the $\CN2$ permits also to quantify the extent of the {\it gray zone}. We note, firstly, that H$_{max}$ = r/$\theta$ (where r is the residual wavefront coherence size after GLAO correction i.e. the residual FWHM) depends on the wavelength.  Using the $\CN2$ composite profiles extended on 20 km as an input (Table \ref{comp_h_gt_1km} and Table \ref{comp_h_lt_1km_corr}), with the combination 'good'-'good', 'typical'-'typical' and 'bad'-'bad', in the boundary layer and in the free atmosphere, the ARGOS simulations provide the residual values r (L. Busoni, private communication). From these values of r that we reported in Table \ref{glao_tab} we can retrieve the values of H$_{max}$ for the wavelengths in the near-infrared range: J, H and K (Table \ref{glao_tab}). Knowing that H$_{min}$ = 200 m, as explained previously, we deduce that the {\it 'gray zone'} extends in the (200 m - 378 m) range and assume its smallest value when we observe in J band and we consider the 'bad'-'bad' case (75\% case). It extends in the (200 m - 3777 m) range and assumes its largest value when we observe in K band and we consider the 'good'-'good' case (25\% case). Fig. \ref{glao} shows the H$_{max}$ (blue, red and yellow lines) for different field of view in the case the residual FWHM of GLAO simulations has been obtained with the central column of Table \ref{comp_h_gt_1km} and Table \ref{comp_h_lt_1km_corr} i.e (case 50$\%$).

\section{OPTICAL TURBULENCE VERTICAL DISTRIBUTION: $\CN2$}
\label{cn2}

\subsection{GS: vertical distribution on the whole troposphere}

\begin{figure*}
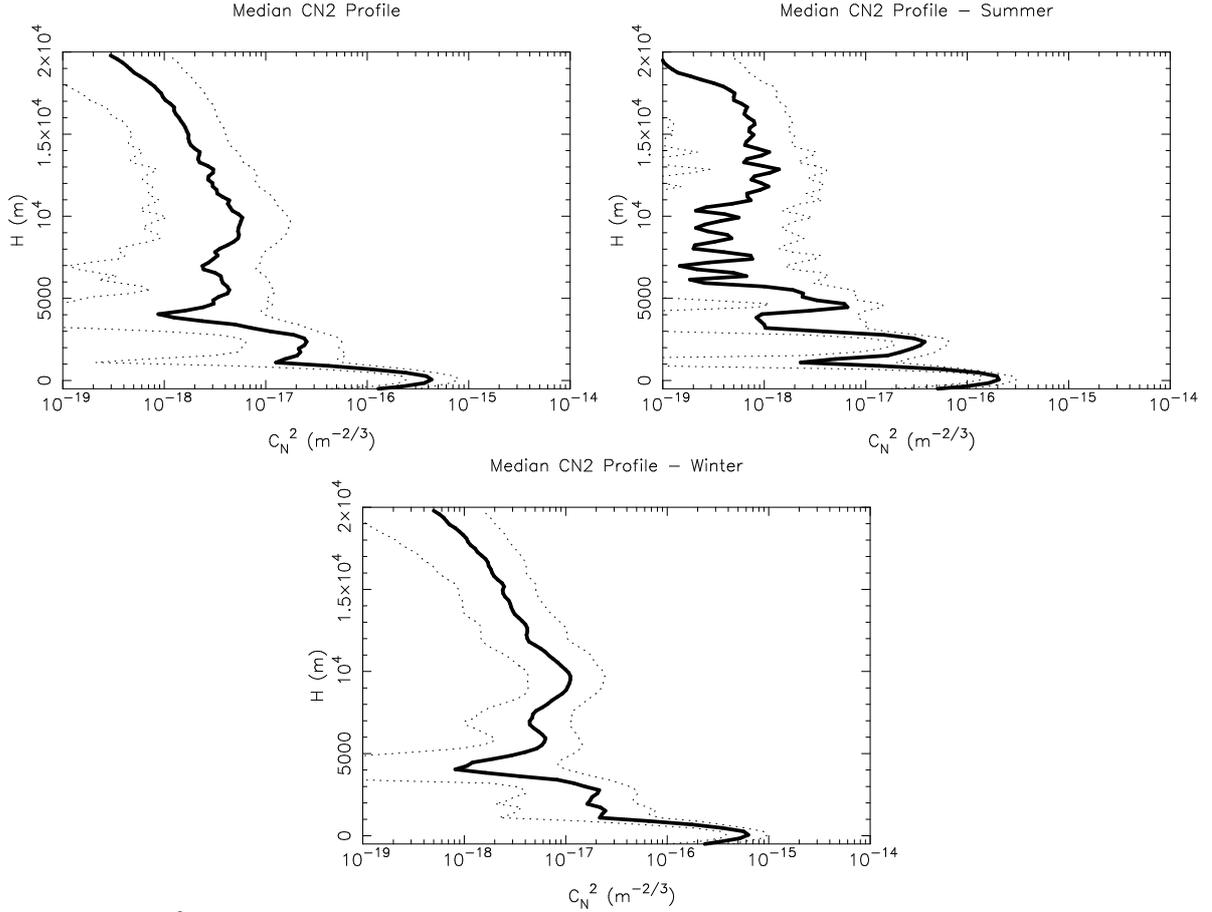

\begin{center}
\includegraphics[width=6cm,angle=-90]{masciadri_art1_fig2a}
\includegraphics[width=6cm,angle=-90]{masciadri_art1_fig2b}
\includegraphics[width=6cm,angle=-90]{masciadri_art1_fig2c}
\caption{Median $\CN2$ profile obtained with the complete sample of 43 nights, the summer [April-June] and winter [October-March] time samples. Results are obtained with the standard GS technique.
\label{cn2_median}}
\end{center} 
\end{figure*}

Figure \ref{cn2_median} shows the median $\CN2$ profile obtained with the whole data-set of 43 nights, the summer and the winter periods. The morphology of the vertical distribution of the optical turbulence ($\CN2$ profile) shows that the greatest turbulence contribution develops in the first kilometer above the ground. Between 1 and 10 km we observe a set of minor peaks changing their position and strength during the year. At around 10 km we observe the typical secondary $\CN2$ peak developed at the jet-stream level. 
For what concerns the seasonal variation we observe that the ground layer bump, responsible for most of the turbulence budget, shows a clear seasonal trend indicating larger turbulence strength in winter than in the summer period. In the free atmosphere we observe the interesting effect of the secondary $\CN2$ peak located at 10 km in winter time that shifts to higher heights ($\sim$ 14 km) and is characterized by a weaker strength in summer time. The latter effect (called {\it '$\alpha$ effect'}) has been put in evidence the first time by Masciadri \& Egner (2006)\cite{MasciadriEgner2006} and widely discussed later in Masciadri et al. (2010)\cite{Masciadri2010}. We refer the reader to the latter paper for a detailed discussion of the seasonal variation trend in the free atmosphere and the {\it '$\alpha$ effect'}.

\subsection{HVR-GS: vertical distribution for h $\le$ 1 km}

The optical turbulence  vertical distribution with high resolution (20-30 m) in the first kilometer is obtained with the method called HVR-GS presented in Egner \& Masciadri (2007)\cite{EgnerMasciadri2007}. The HVR-GS data-set is obtained taking the integral of the $\CN2$ profiles retrieved from the AC frames and redistributing the energy in the first kilometer according to the detected triplets in the CC frames.
Three different strategies can be used to study the turbulence spatial distribution in the boundary layer. The usefulness of each method depends on the application one intends to give to the analysis. We study:\newline\newline
(A)  the median of the $\CN2$ (and/or J) profiles.\newline
(B)  the average of the $\CN2$ (and/or J) profiles. This method is very useful for comparisons of measurements and simulations obtained with atmospherical models. Moreover this operator has the advantage that the mean of J$_{i}$ is equal to the J retrieved from the mean of the C$_{N,i}^{2}$ (where J$_{i}$ and C$_{N,i}^{2}$ refer to the each individual profile). It is not the case for the median.  \newline
(C) the composite profiles as calculated in Section \ref{composit}. This method is very useful for applications to Adaptive Optics.\newline

\begin{figure*}
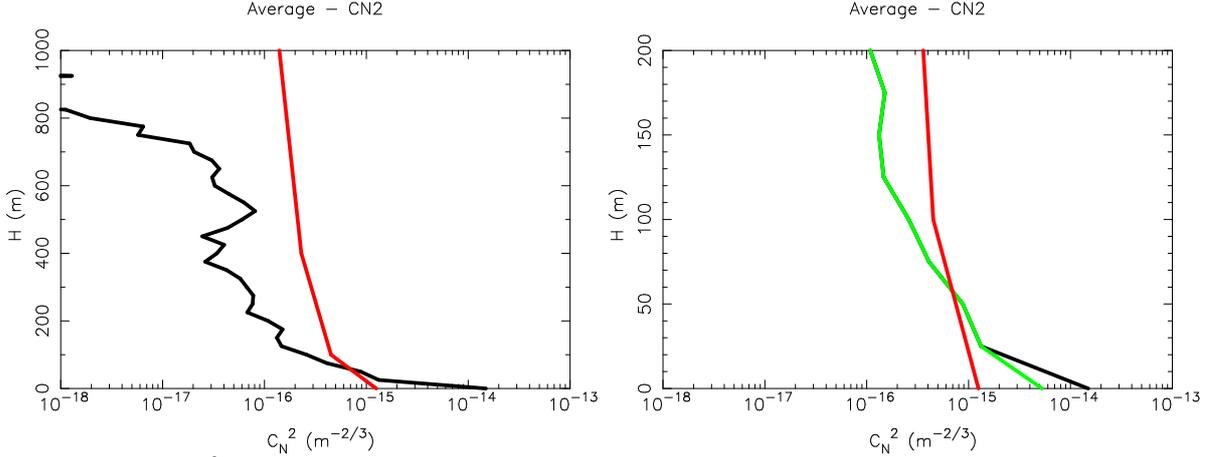

\begin{center}
\includegraphics[width=6cm,angle=-90]{masciadri_art1_fig3a}
\includegraphics[width=6cm,angle=-90]{masciadri_art1_fig3b}
\caption{{\bf Left:} Mean $\CN2$ profile ($\Delta h$$\sim$20-30 m) calculated with the sample 'HVR-GS' and the dome seeing contribution included (black line). Hufnagel model (red line). {\bf Right:} Zoom of the first 200 m. Near the ground it is put in evidence  the different $\CN2$ shape in case the dome contribution is included (black line) and excluded (green line). 
\label{hvr-gs}} 
\end{center}
\end{figure*}

Following the strategy (A) we find that $\CN2$ = 0 for h $>$ 25-30 m. This can be explained with the fact that each vector ($\CN2$ profile) has many zeros. This is due to the fact that, during the monitoring, the system detects spikes and thin layers that changes position and duration with the time. In other words, the use of the median in these cases, can be misleading and it can provide very low values of the $\CN2$. Considering that the strategy (A) is not really useful to characterize these measurements in our case, the strategies (B) and (C) have been used. 
Figure \ref{hvr-gs}-left shows the result obtained following the strategy (B) i.e. the mean of the $\CN2$ profiles ($\Delta$h $\sim$ 25-30 m) calculated from the sample 'HVR-GS'. It is therefore representative of 43 nights. The two profiles (with and without dome contribution) are shown in proximity of the ground (Figure \ref{hvr-gs}-right). To retrieve the typical scale height B of the exponential decay of the mean $\CN2$ profile the measurements done below 125 m have been fitted with an exponential law (Eq.(\ref{eqn:exp})) as we already did in Stoesz et al. (2008)\cite{Stoesz2008}:

\begin{equation}
y=A \cdot e^{(-h/B)}
\label{eqn:exp}
\end{equation}

where A and B are free parameters. The calculation is obviously shown only in the case in which the dome contribution is subtracted. The fit gives A $=$ 3.34$\cdot$10$^{-15}$ and B $=$ 37.4 m (Fig.\ref{hvr-gs}-left). If we limit the analytical fit to the first 30 meters, the scale height B = 28 m. The exponential decay is however an absolutely arbitrary analytical law. The important issue is that these results definitely indicate that the HVR-GS technique is able to put in evidence that the turbulence decays above typical astronomical sites in stable night time conditions in a much sharper way than what has been predicted and quantified in the past. Indeed, the Hufnagel model in proximity of the surface\cite{Roddier1981} in night conditions states that the turbulence scales as h$^{-2/3}$ (Fig.\ref{hvr-gs}). For the Hufnagel model (Fig.\ref{hvr-gs}-left)  the $\CN2$ decreases of one order of magnitude within 1 km while our results indicate that the $\CN2$ decreases of one order of magnitude within $\sim$ 60-70 m (Fig.\ref{hvr-gs}-right). It remains interesting the (800 m - 1 km) range in which visibly a weak turbulence develops. One should expect a smoother connection between the boundary layer and the free atmosphere. We think that this is just an artifact effect due to the use of a quite different resolution below and above 1 km. From the point of view of the AO simulations this small gap should not cause any problems. It should be enough to implement a weak convolution to slightly smooth out the $\CN2$ vertical profiles at the interface located at 1 km so to obtain a less abrupt connection of the $\CN2$ profile above and below 1 km. 

We note that the presence of no signal (therefore $\CN2$= 0) might potentially indicate that the turbulence strength is not equal to zero but simply weaker than the threshold $\CN2$$=$10$^{-16}$ (associated to an equivalent J$=$2.5$\cdot$10$^{-15}$).  
In Masciadri et al. (2009)\cite{Masciadri2009} it has been calculated, in a post-processing phase, the most conservative case in which we assigned $\CN2$$=$10$^{-16}$ where there is no signal. After a more careful investigation we observed that the latest distribution is associated to a too large total J in the boundary layer (the turbulent energy should not be conserved) and, for this reason, it can be discarded. 

\begin{figure*}
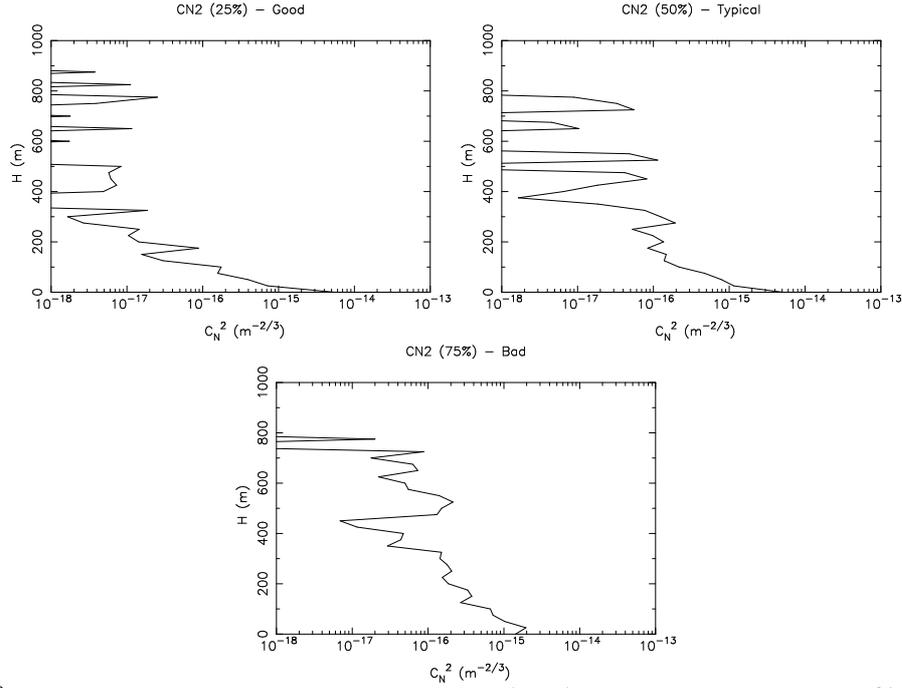

\begin{center}
\includegraphics[width=4.5cm,angle=-90]{masciadri_art1_fig4a}
\includegraphics[width=4.5cm,angle=-90]{masciadri_art1_fig4b}
\includegraphics[width=4.5cm,angle=-90]{masciadri_art1_fig4c}
\caption{Mean $\CN2$ profiles calculated from the corresponding J (or r$_{0}$) values related to the 20-30 $\%$, 45-55 $\%$ and 70-80 $\%$ ranges of the J cumulative distribution. The dome contribution is excluded. In Annex C the numerical values for the correspondent J values are reported.} 
\label{cn2_25_50_75}
\end{center}
\end{figure*}

Figure \ref{cn2_25_50_75} shows the results obtained following the strategy (C) that is the mean $\CN2$ profiles associated to the J (or r$_{0}$) related to the 20-30 $\%$, 45-55 $\%$ and 70-80 $\%$ of the cumulative distribution. In Masciadri et al. (2010)\cite{Masciadri2010} the numerical values for the correspondent J values are reported. Curiously the first grid point near the ground of the 75$\%$ case distribution shows a weaker value with respect to the 25$\%$ and 50$\%$ cases. This is due to the fact that the third quartile dome seeing ($\varepsilon_{d,75}$), that has been subtracted from the original value is particularly large (0.70 arcsec). The morphology of the $\CN2$ in the first kilometer is very interesting showing several thin layers and a very weak turbulence between 800 m and 1 km. We note that the sample on which we calculate the average in each slot [20-30]$\%$, [45-55]$\%$ and [70-80]$\%$ is of a few hundreds of $\CN2$ profiles therefore there are no doubts that this structure reproduces some real distribution. In terms of morphology of the turbulence profile we find therefore that the higher the vertical resolution the thinner the size of the detected layers. This conclusion is perfectly coherent with the turbulence structure resolved by balloons equipped for the $\CN2$ measurements\cite{AzouitVernin2005}. 

\begin{figure*}
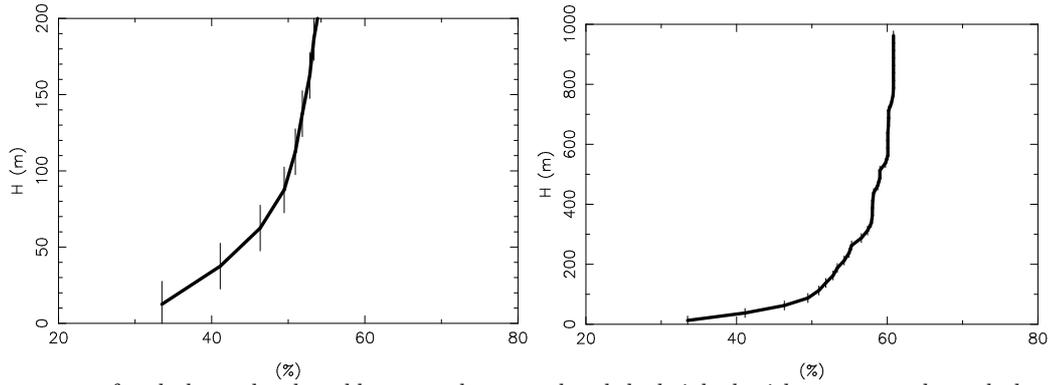

\begin{center}
\includegraphics[width=5cm,angle=-90]{masciadri_art1_fig5a}
\includegraphics[width=5cm,angle=-90]{masciadri_art1_fig5b}
\caption{Percentage of turbulence developed between the ground and the height $h$ with respect to the turbulence developed in the whole atmosphere ($\sim$ 20 km) as retrieved from the HVR-GS measurements and extended to the first kilometer. On the left hand is shown a zoom of the picture centred on the first hundreds of meters.} 
\label{perc}
\end{center}
\end{figure*}

Figure \ref{perc} shows that the percentage of turbulence P(h) developed in the (0, h) range with h in the (0, 1 km) vertical slab. The function P(h) is defined as: 
\begin{equation}
P(h) = \left( {\frac{{\int\limits_0^h {C_N^2 (h^* )dh^* } }}
{{\int\limits_0^{\infty} {C_N^2 (h^* )dh^* } }}} \right) \times 100
\label{perc_eq}
\end{equation}

where the $\CN2$ profile is that associated to the 45-55 $\%$ case (Fig. \ref{cn2_25_50_75}-centre). It is worth to note that the error bars for the HVR-GS technique is of the same order of half of the vertical resolution i.e. $\pm$ 12-15 m. This derives mainly by the definition of the zero point i.e. the ground that is characterized by an uncertainty equivalent to half of the vertical resolution. 

From Fig.\ref{perc} we retrieve that around 50$\%$ of the turbulence developed in the whole 20 km is concentrated below 80$\pm$15m.  This result is substantially different from the preliminary indications obtained in Egner \& Masciadri (2007)\cite{EgnerMasciadri2007} and the turbulence seems to be much more concentrated near the ground. We also note that the $\CN2$ morphology below 1 km decreases in a different way if we look at the sample 'WB' and the 'HVR-GS'. This is not a contradiction and it can be explained with the fact that the turbulence spatial distribution for the 'WB' and the 'HVR-GS' samples is not necessarily the same because the first is based on the AC frames while the second is based on the CC frames. It is interesting to note that the higher the vertical resolution, the sharper is the exponential decay of the morphologic turbulence structure. 

\begin{figure}
\begin{center}
\includegraphics[width=6cm,angle=-90]{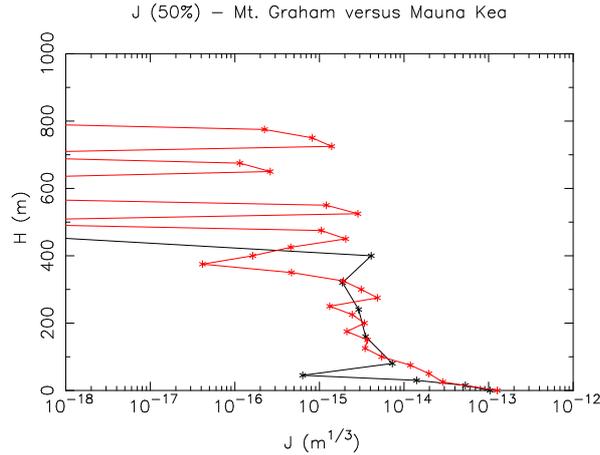}
\caption{Red line: turbulence vertical distribution (J-profile, 50$\%$ case) above Mt. Graham as calculated in Fig.\ref{cn2_25_50_75}-centre. Black line: turbulence vertical distribution (J-profile, 50$\%$ case) above Mauna Kea as calculated by Chun et al. (2009) in Table 6-centre. Mauna Kea measurements extend up to 650 m.} 
\label{mg_vs_mk}
\end{center}
\end{figure}

We note that, simultaneously to our study, some other authors\cite{Chun2009} recently investigated the turbulence structure near the ground at high vertical resolution. Even if the instruments employed were different these studies present many similarities in the results. From a qualitative point of view, we note that, also in that case, the turbulence appears well confined near the surface. From a quantitative point of view, things are more delicate. Looking at Table 3 (in that paper) it appears that their data-reduction is more similar to our method (C) than to the others methods. Figure \ref{mg_vs_mk} shows the J-profile retrieved from Fig. \ref{cn2_25_50_75} (50$\%$ case) overlapped to the J-profile retrieved from Table 3-Chun et al. (2009)\cite{Chun2009}. We note that above Mt. Graham the turbulence vertical sampling is 25 m, above Mauna Kea  the sampling increases from 15 m up to 80 m and it extends only up to 650 m. The turbulence vertical distribution appears very similar. Above Mauna Kea a local minimum is present at $\sim$ 45 m above the ground more or less in correspondence of the abrupt detection break due to the sensitivity threshold from LOLAS (Table 3-Chun et al. (2009)\cite{Chun2009}).  We highlight that the evident  huge {\it 'turbulent vacuum zone'}  between 560 m and 1 km (SLODAR) simply means that the turbulence is not measured in this vertical slab, not that turbulence is not present.

\section{CONCLUSIONS}
\label{concl}
In this paper we present the results of a study aiming to characterize the optical turbulence at Mt. Graham. We present a general overview of the statistics (43 nights) of the $\CN2$ profiles and all the main integrated astroclimatic parameters and their seasonal trends. The main conclusions we achieved are: \newline
{\bf (1)} With a median seeing $\varepsilon$$=$ 0.95 arcsec ($\varepsilon$$=$ 0.72 arcsec without dome contribution), isoplanatic angle $\theta_0$$=$ 2.5 arcsec and a wavefront coherence time $\tau_0$$=$4.8 msec, Mt. Graham confirms its good quality in terms of turbulence characteristics typical of the best astronomical sites in the world. All the integrated astroclimatic parameters (the seeing, the isoplanatic angle, the wavefront coherence time and the equivalent wind speed) show a clear seasonal trend that indicates better turbulence conditions and weaker equivalent wind speed V$_{0}$ in summer with respect to the winter. \newline
{\bf (2)} The ground layer is characterized for the first time with a high resolution (200-250 m and 20-30 m). The turbulence exponentially decays above Mt. Graham with a much sharper profile than what has been supposed so far and expressed with the Hufnagel model. Three different strategies of analysis aiming to investigate the morphology of the turbulence spatial distribution have been presented. We find that around 50$\% $ of the turbulence developed in the whole atmosphere is concentrated below 80$\pm$15 m from the ground and 60$\%$ of the turbulence in the first kilometer. This evidence together with the favorable large $\theta_{0}$ observed above Mt. Graham (particularly in the spring/summer time) represent extremely favorable conditions for astronomical observations assisted by a LGS/GLAO system such as ARGOS. \newline
{\bf (3)} We observe that the higher is the vertical resolution of the tool used to measured the turbulence vertical distribution the sharper is the turbulence decreasing. \newline
{\bf (4)} The percentage of turbulence developed below the primary mirror of the LBT (i.e. $\sim$ 35 m from the ground) is around 33$\%$. However this estimate has to be considered with precaution because the uncertainty (2$\sigma$$\sim$ 25-30 m) is of the same order of magnitude of the vertical resolution ($\Delta$H $\sim$ 25-30 m) and in the first hundred meters the turbulence decreases very sharply.\newline
{\bf (5)} It appears evident that at Mt. Graham the turbulence decreases above the ground similarly to what observed above Mauna Kea in a more or less simultaneous study (Chun et al. 2009)\cite{Chun2009} performed with different instrumentation. \newline
{\bf (6)} A composite distribution of the turbulence on the whole 20 km is calculated to be used as input of AO simulations of the LBT Laser Guide Star system named ARGOS and the calculation of the gray zones for the near-infrared J, H and K band is done. The gray zone extends from a minimum of (200m - 378 m) in J band with the 'bad'-'bad' case up to a maximum of (200m - 3777 m) in K band in the case 'good'-'good' case. \newline
{\bf (7)} A clear $\CN2$ seasonal variation trend has been observed in proximity of the ground and  in the jet-stream regions. These measurements confirm the first evidence of the $\CN2$ seasonal trend observed by Masciadri \& Egner (2006)\cite{MasciadriEgner2006} above other astronomical site. The physical model proposed by Masciadri \& Egner (2006)\cite{MasciadriEgner2006} that is able to explain the seasonal effect of the secondary peak of the $\CN2$ called {\it  '$\alpha$ effect'}, is confirmed and refined.\newline
{\bf (8)} For the first time we observed a seasonal trend of the dome seeing. This is certainly a topic that deserves a more careful investigation in the future.  \newline

\acknowledgments     
 
This study has been funded by the Marie Curie Excellence Grant (FOROT) - MEXT-CT-2005-023878. 
We sincerely thank the VATT and LBT staff for the support offered during the Generalized Scidar runs at Mt. Graham. ECMWF products are extracted by the MARS catalog $http://www.ecmwf.int$ and authors are authorized to use them by the Meteorologic Service of the Italian Air Force. 



\end{document}